\documentclass[journal,10pt]{IEEEtran}
\usepackage{amsfonts}
\IEEEoverridecommandlockouts

\ifCLASSINFOpdf
\else
\fi
\usepackage{epsfig}
\usepackage{graphicx}
\usepackage{subfigure}
\usepackage{psfig}
\usepackage{epsf}
\usepackage[cmex10]{amsmath}
\usepackage{booktabs}
\usepackage{fancyhdr}
\usepackage{stfloats}
\usepackage{color}

\begin{document}
 \title{Resource Allocation for a Wireless Powered Integrated Radar and Communication System}
\author{\IEEEauthorblockN{Yifan Zhou, Huilin Zhou, Fuhui Zhou, Yongpeng Wu, \emph{Senior Member, IEEE}, Victor C. M. Leung, \emph{Fellow, IEEE}}

\thanks{The research was supported by the National Natural Science Foundation of China (61701214, 61561034, and 61701301), the Excellent Youth Foundation of Jiangxi Province (2018ACB21012), the Young Natural Science Foundation of Jiangxi Province (20171BAB212002), the Postdoctoral Science Foundation (2017M610400, 2017KY04, 2017RC17), and Young Elite Scientist Sponsorship Program by CAST.

Y. Zhou, H. Zhou and F. Zhou are with the School of Information Engineering, Nanchang University,  China. (e-mail: zhouyifan@email.ncu.edu.cn, zhouhuilin@ncu.edu.cn, zhoufuhui@ieee.org). F. Zhou is also with  the Postdoctoral Research Station of Environment Science and Engineering, Nanchang University. Y. Wu is with with Shanghai Key Laboratory of Navigation and Location Based Services, Shanghai Jiao Tong University, Minhang, China (email: yongpeng.wu2016@gmail.com). V. C. M. Leung is with the Department of Electrical and Computer Engineering, the University of British Columbia, Vancouver, BC, V6T 1Z4, Canada (email: vleung@ece.ubc.ca). The corresponding author is F. Zhou.}}
\maketitle
\begin{abstract}
The integrated radar and communication system is promising in the next generation wireless communication networks. However, its performance is confined by the limited energy. In order to overcome it, a wireless powered integrated radar and communication system is proposed. An energy minimization problem is formulated subject to constraints on the  radar and communication performances. The energy beamforming and radar-communication waveform are jointly optimized to minimize the consumption energy.  The challenging non-convex problem is solved by using semidefinite relaxation and  auxiliary variable methods. It is proved that the optimal solution can be obtained. Simulation results demonstrate that our proposed optimal design outperforms the benchmark scheme.
\end{abstract}
\IEEEpeerreviewmaketitle
\section{Introduction}
Integrated radar and communication (IRC) systems are promising in the next generation wireless communication networks since radar and communication functions can be simultaneously performed at a common frequency band in a single platform and  the antenna and signal processing hardware can be shared \cite{G.C. Tavik}.  Recently, it has inspired widely investigations from industry and academia, such as the development of the electronic warfare and intelligent transportation system \cite{Jaber Moghaddasi}. In the IRC system, it is of crucial importance to design waveform for improving the radar and communication performance. Up to now, there have been two main categories of waveform design, namely, multiplexing waveform \cite{Aboulnasr Hassanien} and identical waveform \cite{Cenk Sahin}. For the first one, the radar and communication waveforms are multiplex. However, the resource utilization efficiency is low. In contrast, for the identical waveform, the resource efficiency is high since the waveform is shared by radar and communication functions. Thus, we focus on the second one in this paper.

Since the identical waveform relies on the traditional communication waveform, and orthogonal frequency division multiplexing (OFDM) waveform has high spectrum efficiency (SE), high   modulation flexibility and strong tolerance for inter-symbol interference, there are many literatures devoted to exploiting OFDM to simultaneously perform radar and communication functions for improving SE \cite{Christian Sturm}-\cite{Chenguang Shi}. In \cite{Christian Sturm}, the authors focused on signal processing in the IRC system. However, the OFDM waveform has not been optimized. The authors in \cite{Yongjun Liu} studied the optimization of OFDM waveform and designed flexible waveform. The designed waveform is adaptive to the environment and can improve the performance. Recently, the authors in \cite{Chenguang Shi1} and  \cite{Chenguang Shi} studied the resource allocation problem and  the robust OFDM radar waveform design problem, respectively. The optimal power allocation strategy and the robust waveform have been designed.

Although the performance of the IRC system can be improved by the designed waveform in \cite{Christian Sturm}-\cite{Chenguang Shi}, it is still confined by the finite energy of devices. Fortunately, wireless power transfer (WPT) has been proposed to prolong the operation time of devices \cite{Rui Zhang}.  Low-power devices can harvest energy from the wireless environment. In \cite{J. Xu}, it was shown that WPT can improve the energy efficiency (EE) of wireless communication networks. Thus, in order to overcome the finite operational energy problem and improve the performance of IRC systems, a wireless powered IRC system is studied in this paper. The transmitter exploits the collected energy from a wireless power station to simultaneously perform radar and communication functions. Moreover, a protocol called  \lq\lq harvest-then-transmit\rq\rq \  is applied in order to facilitate the implementation of the wireless powered IRC system \cite{Derrick Wing Kwan Ng}.

Different from the works in \cite{Chenguang Shi1} and  \cite{Chenguang Shi}, it is the first time that a wireless powered IRC system is proposed. The transmission energy of the wireless power station is minimized by jointly optimizing the energy beamforming vector, energy harvesting time and the transmit power. It is proved that the optimal solution can be obtained. Then, by demonstrating that the optimal beamforming matrix after semi-definition relaxation (SDR) is rank-one, we prove that the constraints of the non-convex problem are always achievable. Simulation results show that our proposed optimal design outperforms the benchmark scheme.

\emph{Notations}: Boldface capital letters represent matrixes; boldface lower case letters represent vectors. $\textbf{I}$ and $\textbf{0}$ denote the identity matrix and the matrix with zero entries (their size is determined from context), respectively. $\textbf{x}^{H} $ represents the conjugate transpose of a vector $\textbf{x}$. $\mathbb{E}{\{{\cdot}\}}$ represents the expectation operator. $\text{Tr}{\{{\cdot}\}}$ is the trace of the square matrix argument. $\textbf{A}{\succeq}{\textbf{0}}$ means $\textbf{A}$ is a Hermitian positive semi-definite matrix. $\text{Rank}(\textbf{A}) $ is the rank of matrix $\textbf{A}$. ${\mathbb{C}^{M{\times}N}}$ represents a $M$-by-$N$ dimensional complex matrix set. $\text{H}_+^{N}$ denotes the set of all $N$-by-$N$ Hermitian positive semi-definite matrices. $\mathbb{R}_+$ denotes the set of all nonnegative real numbers.

\section{System Model}
\begin{figure}[!t]
\centering
\includegraphics[width=3.0 in]{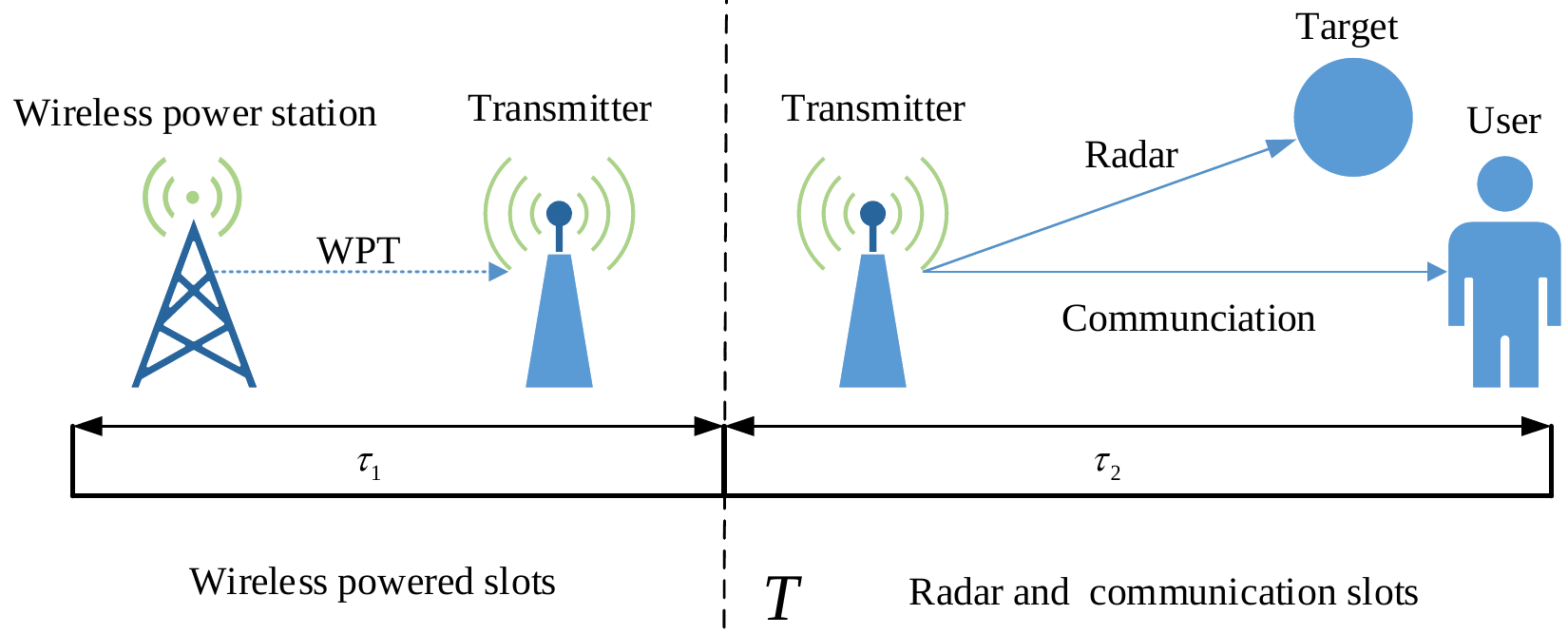}
\caption{The wireless powered IRC system model.}
\label{fig.1}
\end{figure}
In this section, as shown in Fig. 1, in order to overcome the problem that the performance of the IRC system is limited by the operational time due to its finite energy, a wireless powered IRC system is studied. This system has great potential to be applied in those scenarios that the transmitter of the IRC system cannot be wired charged or the cost for achieving wired charging is extremely high \cite{X. Lu}.  A \lq\lq harvest-then-transmit\rq\rq \ protocol is applied. It consists of two phases. In the WPT phase, a wireless power station transmits energy to a transmitter. In the IRC phase, the transmitter simultaneously performs radar and communication functions. The wireless powered time slot is denoted by $\tau_1$; the radar and communication slot is denoted by $\tau_2$, and the total time is denoted by $T_s$.

In the WPT phase, the wireless power station transmits the energy RF signal to the transmitter and the transmitter harvests energy. It is assumed that the wireless power station is equipped with $N_t$ antennas. The transmitter and all users have one antenna. The RF signal received by the transmitter from the wireless power station can be expressed as
\begin{align}
&{y_{T}}={\textbf{h}^{H}}{\textbf{x}}+{n},
\end{align}
where ${\textbf{h}\!\in\!{\mathbb{C}^{N_t\times1}}}$ denotes the channel gain vector from the wireless power station to the transmitter; ${\textbf{x}\!\in\!{\mathbb{C}^{N_t\times1}}}$ is the signal vector that satisfies ${\mathbb{E}}{\{{\textbf{x}^{H}{\textbf{x}}}\}=P}$, and ${\textbf{x}}\!=\!{\textbf{w}_e}s$; $s$ is RF energy signal with unit power; $P$ is the transmit power of the wireless power station with its unit being W; ${\textbf{w}_e}\!\in\!\mathbb{C}^{N_t\times1}$ is the energy beamforming vector, and $n$ is the additive white Gaussian noise (AWGN) with zero mean.
Thus, the harvested energy of the transmitter can be given as
\begin{align}
&{E_1}={\eta}{\tau_1}{|{{\textbf{h}^{H}}{\textbf{w}_e}}|^2},
\end{align}
where $\eta\in[0,1]$ represents the energy conversion efficiency of the energy harvesting circuit.

In the IRC phase, $N_s$ consecutive integrated OFDM signals are transmitted during the time slot $\tau_2$, given as \cite{Yongjun Liu}
\begin{align} \notag
{z(t)}=&{e^{j2{\pi}{f_c}t}}{\sum\limits_{n=0}^{{N_s}-1}}{\sum\limits_{m=0}^{{N_c}-1}}{a_m}{c_{m,n}}{e^{j2{\pi}m{\Delta}f(t-nT_s)}}\\
&{\times}{rect[(t-nT_s)/T_s]},
\end{align}
where $f_c$ is the center frequency with its unit being Hz; $N_c$ is the number of subcarrier; $a_m$ is the amplitude of the $m$th subcarrier that needs to be designed;
$c_{m,n}$ is the phase encoding of the $m$th subcarrier at the $n$th OFDM signal; $\Delta{f}$ is the interval between two adjacent OFDM subcarriers in the frequency domain; $T_s$ is the duration of each completed OFDM symbol, and $rect[t]$ is rectangle function, which is equal to one for $0{\le}t{\le}1$; otherwise, it is zero.

For radar target detection and valuation, the conditional mutual information (MI) is an appropriate metric to evaluate the estimation accuracy of the target impulse response. Let $g(t)$ denote the impulse response of an extended target. The signal received by the radar receiver is given as
\begin{align}
&{y(t)}={{\int}_{-\infty}^{+\infty}}{g(\tau)}{z(t-\tau)}{\,}{\mathrm{d}}{\tau}+n(t),
\end{align}
where $n(t)$ is complex AWGN with zero mean and power spectral density $N(f)$. In order to improve the estimation accuracy of the target impulse response, the conditional MI between the received signal and the impulse response should be as large as possible \cite{Yang Yang}. The conditional MI can be expressed as \cite{Yongjun Liu}
\begin{align}
&{I_{t}(y(t);g(t)|z(t))}={\frac{1}{2}}{\Delta}f{T_p}{\sum\limits_{m=0}^{{N_c}-1}}{\log}_{2}(1+{p_m}{v_m}),
\end{align}
where ${p_m}=|{a_m}|^2$ denotes the $m$th subcarrier power  with its unit being W; ${T_p}={N_s}{T_s}$ is the total duration of the transmitted OFDM signal, namely, $T_p=\tau_2$; ${v_m}={N_s}{T_s^2}{|G(f_m)|}^2/{(N(f_m)T_p)}$ can be regarded as the signal to noise ratio (SNR); ${f_m}={f_c}+m{\Delta}f$ is the frequency of the $m$th subcarrier, and $G(f)$ is the Fourier transform of $g(t)$.

For the communication process, the data information rate (DIR) is an important performance metric, and the frequency selective fading channel is considered. Thus, the total DIR is given as
\begin{align}
&{C_t}={\Delta}f{\sum\limits_{m=0}^{{N_c}-1}}{\log}_{2}(1+{p_m}{\varpi}_m),
\end{align}
where ${\varpi}={|{h_m}|^2}/{\sigma}_c^2$ can be interpreted as the SNR; $h_m$ is the channel gain of the $m$th subchannel, and ${\sigma}_c^2$ is the noise power at the user receiver. In order to guarantee the radar and communication performance, the minimum MI and DIR are required to be considered. They can be expressed as
\begin{subequations}
\begin{align}
&{\qquad}{\quad}{C1:} \ {\frac{1}{2}}{\Delta}f{\tau_2}{\sum\limits_{m=0}^{{N_c}-1}}{\log}_{2}(1+{p_m}{v_m}){\geq}{R_r}, \\
&{\qquad}{\quad}{C2:} \ {\Delta}f{\tau_2}{\sum\limits_{m=0}^{{N_c}-1}}{\log}_{2}(1+{p_m}{\varpi}_m){\geq}{R_c},
\end{align}
\end{subequations}
where $R_r$ and $R_c$ are the minimum performance requirements for the radar and communication process, respectively. Moreover, due to the energy harvesting causal constraint, the consumption energy cannot be larger than the harvesting energy, namely,
\begin{align}
&{\qquad}{\quad}{C3:} \ {\tau_2}{\sum\limits_{m=0}^{{N_c}-1}}{p_m}{\leq}{\eta}{\tau_1}{|{\textbf{h}^H}{\textbf{w}_e}|}^2.
\end{align}
\section{Energy Beamforming and Waveform Design}


In this section, an energy beamforming and waveform design problem is formulated. In order to minimize the energy of the wireless power station, the optimization problem is formulated as $\text{P}_0$, given as
\begin{subequations}
\begin{align}
&\text{P}_0:\ \min_{{\textbf{w}_e},{\boldsymbol{\tau}},p_m} {\tau_1}{\text{Tr}}{({\textbf{w}_e}{\textbf{w}_e^H})} \\
&\text{s.t.} {\quad}\ \ \, {C1-C3},\\
&{\qquad}{\quad} {C4:}\ {\text{Tr}}{({\textbf{w}_e}{\textbf{w}_e^H})}{\leq}P, \\
&{\qquad}{\quad}{C5:} \ {\tau_1}+{\tau_2}{\leq}T,{\quad}{\{{\tau_1},{\tau_2}\}}{\geq}0, \\
&{\qquad}{\quad}{C6:} \ {p_m}{\geq}0,{\quad}m=0,1,{\cdot\cdot\cdot},{N_c}-1,
\end{align}
\end{subequations}
where ${\boldsymbol{\tau}}=[\tau_1,\tau_2]$.  $C4$ is the peak power constraint due to the nonlinearity of power
amplifiers in practice. It is given to protect the transmitter \cite{Fuhui Zhou}. $C5$ is the total time constraint.

$\text{P}_0$ is a non-convex problem due to the existence of couples among different optimization variables, such as ${\textbf{w}_e}$ and $\tau_1$. In order to solve it, several auxiliary variables are used, namely, $\textbf{Q}_e={\textbf{w}_e}{\textbf{w}_e^H}$, ${\overline{\textbf{Q}}_e}={\tau_1}\textbf{Q}_e$, $\gamma_m={\tau_2}p_m, m=0,1,{\cdot\cdot\cdot},{N_c}-1$. Thus, $\text{P}_0$ can be equivalently as $\text{P}_1$, given as
\begin{subequations}
\begin{align}
&\text{P}_1:\ \min_{{\overline{\textbf{Q}}_e},{\boldsymbol{\tau}},\gamma_m} {\text{Tr}}({\overline{\textbf{Q}}_e}) \\
&\text{s.t.} {\quad}\ \ \,{C1:}\ {\frac{1}{2}}{\Delta}f{\tau_2}{\sum\limits_{m=0}^{{N_c}-1}}{\log}_{2}(1+{\frac{{\gamma_m}{v_m}}{\tau_2}}){\geq}{R_r},\\
&{\qquad}{\quad}{C2:} \ {\Delta}f{\tau_2}{\sum\limits_{m=0}^{{N_c}-1}}{\log}_{2}(1+{\frac{{\gamma_m}{{\varpi}_m}}{\tau_2}}){\geq}{R_c}, \\
&{\qquad}{\quad}{C3:} \ {\sum\limits_{m=0}^{{N_c}-1}}{\gamma_m}{\leq}{\eta}{\text{Tr}}({\textbf{h}}{\textbf{h}^H}{\overline{\textbf{Q}}_e}), \\
&{\qquad}{\quad}{C4:} \ {\text{Tr}}{({\overline{\textbf{Q}}_e})}{\leq}{\tau_1}P, \\
&{\qquad}{\quad}{C5:} \ {\tau_1}+{\tau_2}{\leq}T,{\quad}{\{{\tau_1},{\tau_2}\}}{\geq}0, \\
&{\qquad}{\quad}{C6:} \ {\gamma_m}{\geq}0,{\quad}m=0,1,{\cdot\cdot\cdot},{N_c}-1, \\
&{\qquad}{\quad}{C7:} \ {\overline{\textbf{Q}}_e}{\succeq}{\textbf{0}}, \\
&{\qquad}{\quad}{C8:} \ \text{Rank}({\overline{\textbf{Q}}_e})=1,
\end{align}
\end{subequations}
where $C7$ and $C8$ are given to guarantee that ${\overline{\textbf{Q}}_e}$ is a semi-definite and rank-one matrix since the matrix of the form ${\textbf{w}_e}{\textbf{w}_e^H}$ must be semi-definite and rank-one \cite{G. H. Golub}. The problem $\text{P}_1$ is still non-convex due to the constraint $C8$. We use the SDR technique to relax the non-convex optimization problem \cite{Yongwei Huang}. $\text{P}_1$ can be expressed as
\begin{subequations}
\begin{align}
&\text{P}_2:\ \min_{{\overline{\textbf{Q}}_e},{\boldsymbol{\tau}},\gamma_m} {\text{Tr}}({\overline{\textbf{Q}}_e}) \\
&\text{s.t.} {\quad}\ \ \,C1-C7.
\end{align}
\end{subequations}
Note the inequality constraints in $C1$ and $C2$ are convex, since they are the perspective function of $\log_2(1+x)$. Moreover, other constraints are linear. It is easy to prove that $\text{P}_2$ is convex and can be efficiently solved by using the interior-point method \cite{Fuhui Zhou}. Let $({\overline{\textbf{Q}}_e^{opt}},{{\boldsymbol{\tau}}^{opt}},{\gamma_m^{opt}})$ denote the optimal solution of $\text{P}_2$. Based on solving $\text{P}_2$, Theorem 1 can be obtained.

\emph{\textbf{Theorem 1:}} Provided that $\text{P}_2$ is feasible, the optimal solution ${\overline{\textbf{Q}}_e^{opt}}$ always exists and is rank-one.

\emph{\textbf{Proof:}} In order to prove that the optimal solution ${\overline{\textbf{Q}}_e^{opt}}$ is rank-one, the following problem is firstly considered.
\begin{subequations}
\begin{align}
&\text{P}_3:\ \,\min_{{\overline{\textbf{Q}}_e}} \ {\text{Tr}}({\overline{\textbf{Q}}_e}) \\
&\text{s.t.} {\quad}\ {\sum\limits_{m=0}^{{N_c}-1}}{\gamma_m^{opt}}{\leq}{\eta}{\text{Tr}}({\textbf{h}}{\textbf{h}^H}{\overline{\textbf{Q}}_e}), \\
&{\quad}{\quad}\ \, \ {\overline{\textbf{Q}}_e}{\succeq}{\textbf{0}}.
\end{align}
\end{subequations}
Obviously, $\text{P}_3$ is convex and can achieve the optimal solution ${\overline{\textbf{Q}}_e^{\ast}}$. It is seen from eq. (11) and eq. (12) that that ${\overline{\textbf{Q}}_e^{opt}}$ is also a feasible solution of $\text{P}_3$. It indicates that ${\text{Tr}}({\overline{\textbf{Q}}_e^{\ast}}){\leq}{\text{Tr}}({\overline{\textbf{Q}}_e^{opt}})$. Combining the constraints of $\text{P}_2$, we can obtain ${\text{Tr}}({\overline{\textbf{Q}}_e^{\ast}}){\leq}{\tau_1^{opt}}P$. Thus, $({\overline{\textbf{Q}}_e^{\ast}},{{\boldsymbol{\tau}}^{opt}},{\gamma_m^{opt}})$ is also a feasible solution of $\text{P}_2$ and we can obtain ${\text{Tr}}({\overline{\textbf{Q}}_e^{\ast}}){\geq}{\text{Tr}}({\overline{\textbf{Q}}_e^{opt}})$. Hence, we have ${\text{Tr}}({\overline{\textbf{Q}}_e^{\ast}})={\text{Tr}}({\overline{\textbf{Q}}_e^{opt}})$; in other words, $({\overline{\textbf{Q}}_e^{\ast}},{{\boldsymbol{\tau}}^{opt}},{\gamma_m^{opt}})$ is also the optimal solution of $\text{P}_2$. Next, we need to prove that ${\overline{\textbf{Q}}_e^{\ast}}$ satisfies the rank-one property.

For $\text{P}_3$, the Lagrangian function is given as
\begin{align}
&{L[{\overline{\textbf{Q}}_e},{\textbf{Y}},{\mu}]}\!=\!{\text{Tr}}({\overline{\textbf{Q}}_e})\!-\!{\text{Tr}}({\overline{\textbf{Q}}_e}{\textbf{Y}})\!+\!{\mu}[{\rho}-{\text{Tr}}({\textbf{h}}{\textbf{h}^H}{\overline{\textbf{Q}}_e})],
\end{align}
where ${\mu}{\in}{\mathbb{R}_+}$ and ${\textbf{Y}}{\in}{\text{H}_+^{N_t}}$ are dual variables associated with (12b) and (12c), and we define $\rho={\frac{1}{\eta}{\sum_{m=0}^{{N_c}-1}}{\gamma_m^{opt}}}{\geq}0$. The corresponding KKT conditions are
\begin{subequations}
\begin{align}
&{\textbf{Y}}={\textbf{I}}-\mu({\textbf{h}}{\textbf{h}^H}), \\
&{\textbf{Y}}{\overline{\textbf{Q}}_e}={\textbf{0}}, {\qquad}{\overline{\textbf{Q}}_e}{\succeq}{\textbf{0}}.
\end{align}
\end{subequations}
Since the rank of the ${\textbf{h}}{\textbf{h}^H}$ is one, ${\textbf{Y}}$ has at least $N_t-1$ positive eigenvalues, and one has
\begin{align}
&\text{Rank}({\textbf{Y}})=\text{Rank}({\textbf{I}}-\mu({\textbf{h}}{\textbf{h}^H})){\geq}{N_t-1}.
\end{align}
Based on (14), one has that $\text{Rank}({\textbf{Y}})$ is either $N_t$ or $N_t-1$. For $\text{Rank}({\textbf{Y}})=N_t$, ${\overline{\textbf{Q}}_e}$ must be ${\textbf{0}}$. It is contradictory with (12b) in $\text{P}_3$. For $\text{Rank}({\textbf{Y}})=N_t-1$, ${\overline{\textbf{Q}}_e}$ lies in the null space of ${\textbf{Y}}$, whose dimension is one. This means that the optimal solution ${\overline{\textbf{Q}}_e^{opt}}$ must be rank-one.

Since ${\overline{\textbf{Q}}_e^{opt}}$ is rank-one, the optimal solution of $\text{P}_2$ is also the optimal solution of $\text{P}_1$. Moreover, this solution is the globally optimal since $\text{P}_2$ is convex. Furthermore, the optimal solution of $\text{P}_0$, ${\textbf{w}_e^{opt}}$, can be obtained via singular value decomposition.

\section{Simulation Results}
\begin{figure*}[!t]
\centering
\subfigure[The transmission energy  versus the minimum required MI]
{\includegraphics[height=2 in,width=2.3 in,angle=0]{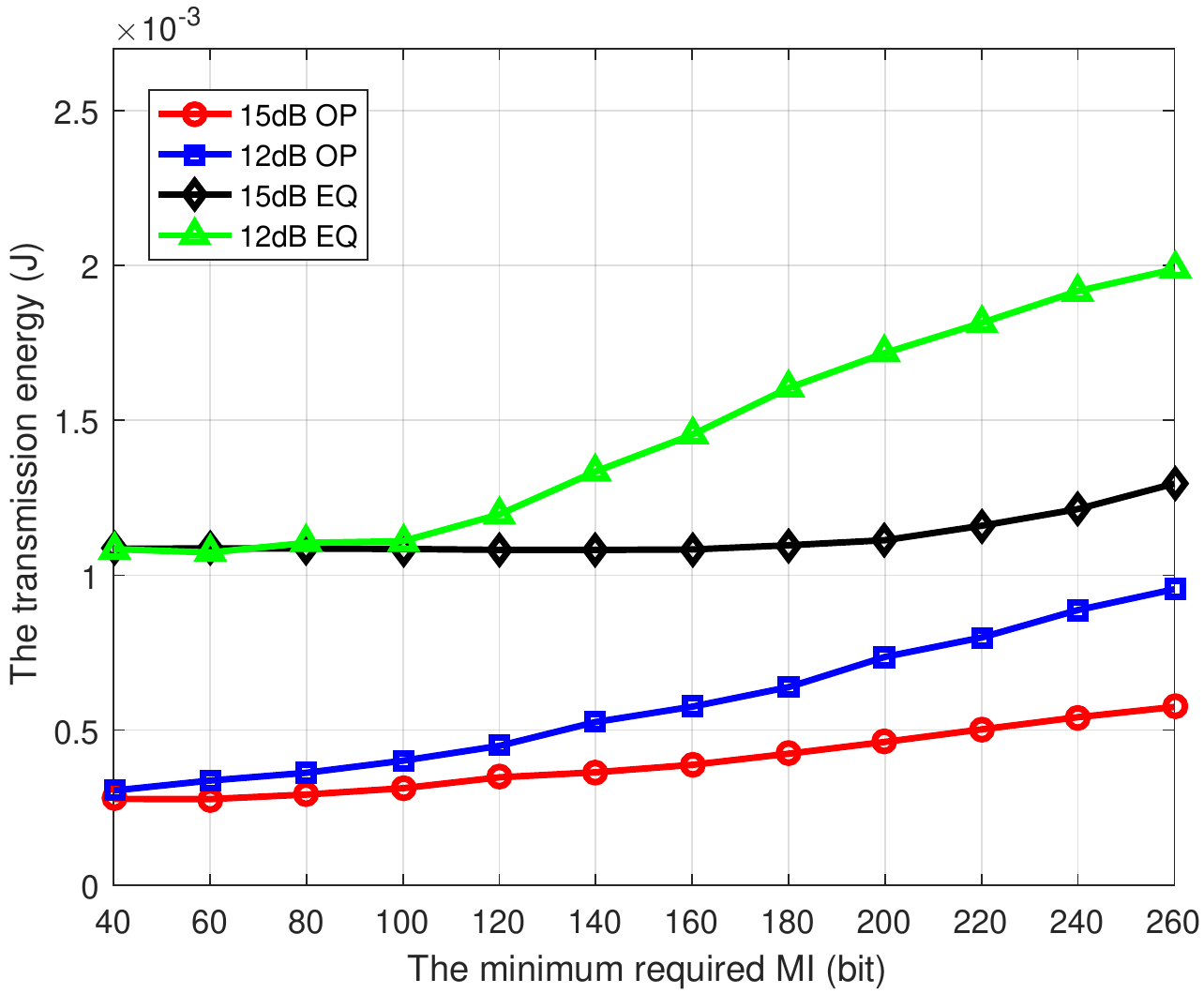}}
\subfigure[The harvesting energy and transmission time versus the minimum required MI]
{\includegraphics[height=2 in,width=2.3 in,angle=0]{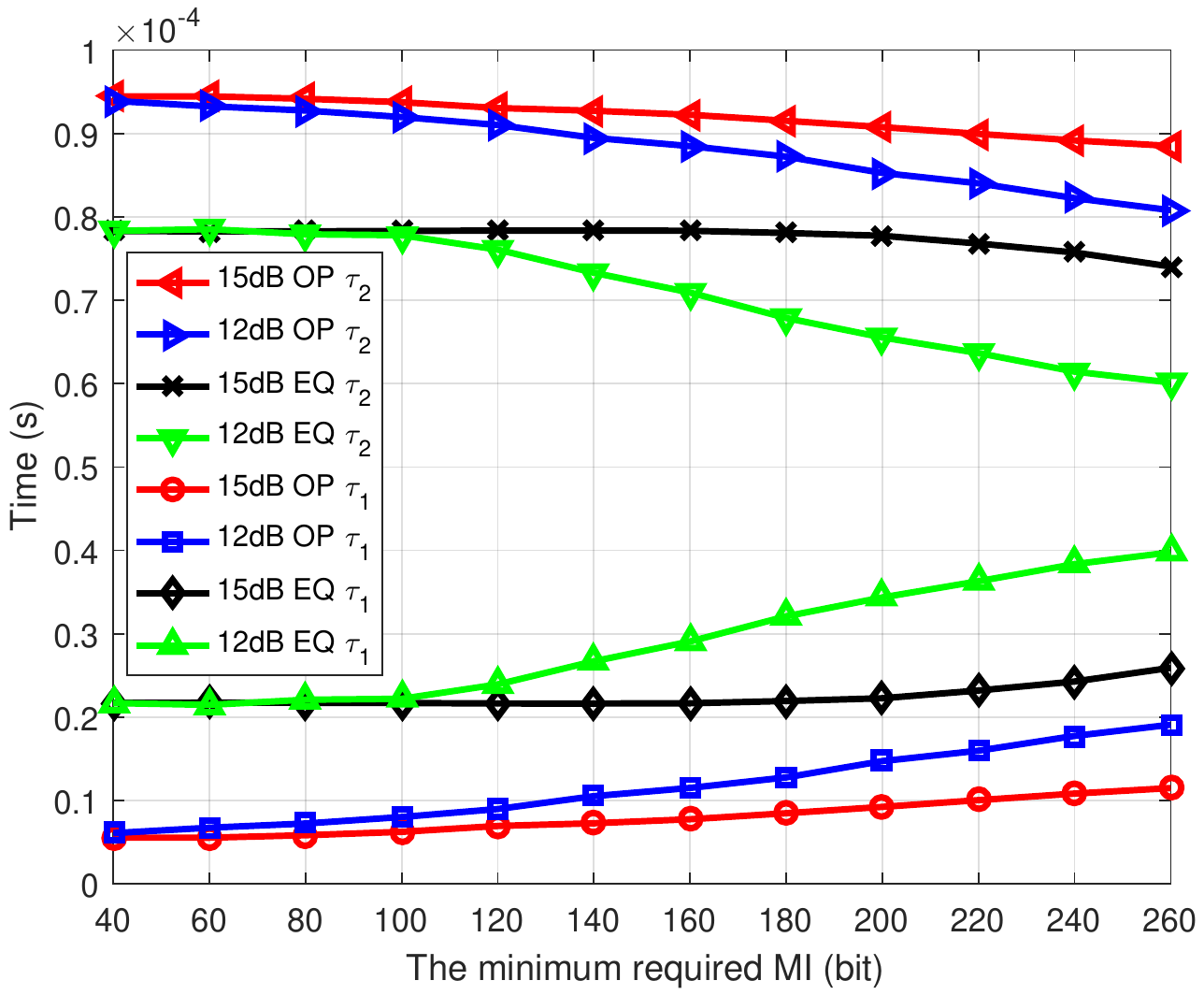}}
\subfigure[The transmission energy  versus the minimum required DIR.]
{\includegraphics[height=2 in,width=2.3 in,angle=0]{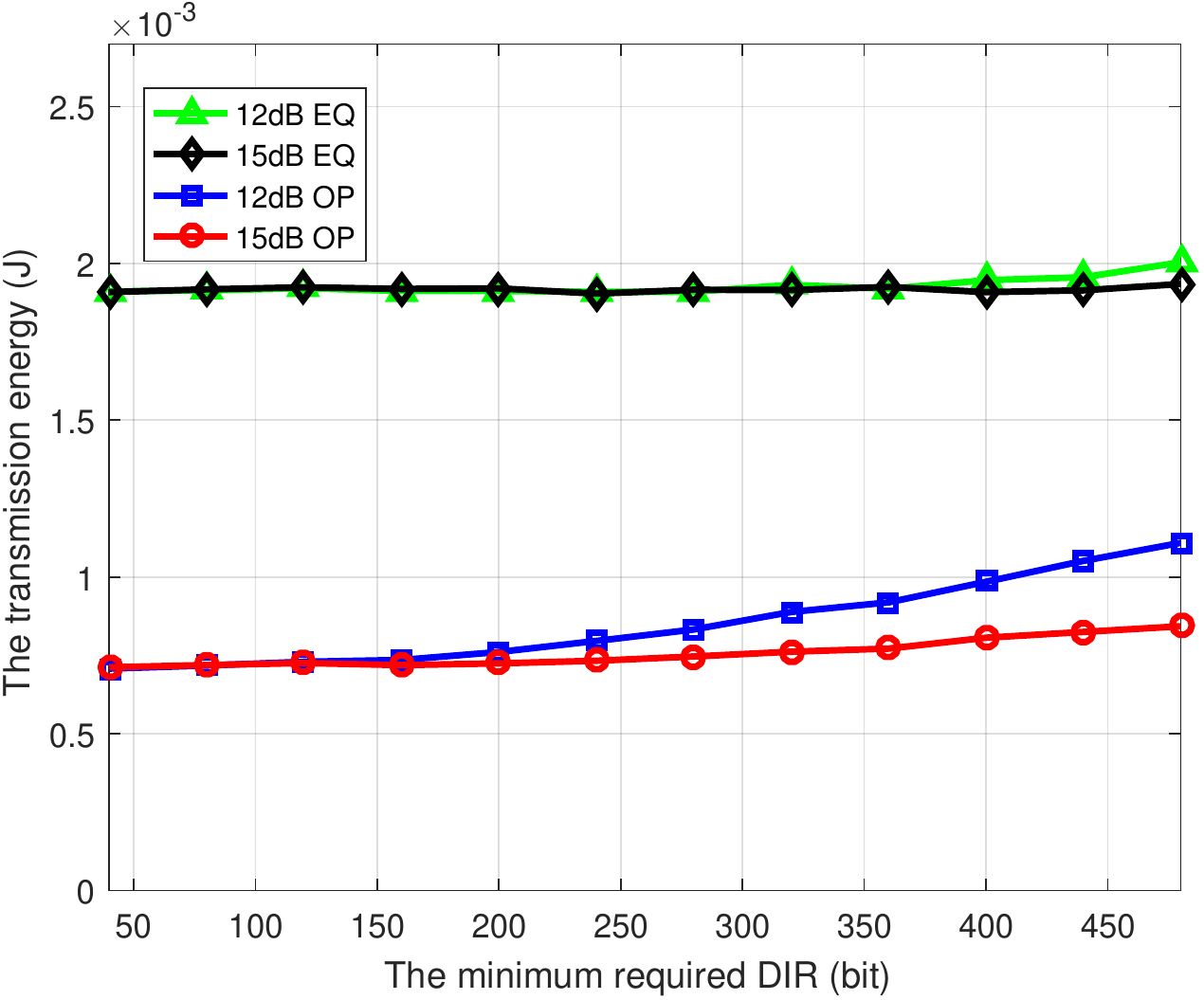}}
\label{fig2}
\end{figure*}

In this section, simulation results are presented to demonstrate the effectiveness of the proposed design. The simulation parameters are based on the work in \cite{Yongjun Liu}, given as: $N_c=128$, ${\Delta}{f}=2.5{\times}{10^{5}}$ Hz, $T_s=5{\times}{10^{-6}}$ s, $T=1{\times}{10^{-4}}$ s, $N_t=5$, $P=50$ W and $\eta$ is 0.5. Both the frequency response of radar target and the frequency response of the communication channels are subject to the standard normal distribution, and the channel coefficients between each antenna of the wireless power station and IRC transmitter follow Rayleigh fading with the distribution ${\cal C}{\cal N}(0,0.2)$.  For a comparison, the benchmark scheme that the transmit power is equally allocated in each sub-channel is applied.

Fig. 2(a)  compares the minimum transmission energy of the wireless power station required by using our proposed scheme with that achieved by using the benchmark scheme. Our proposed scheme is denoted by \lq\lq OP\rq\rq\ while the benchmark scheme is represented by \lq\lq EQ\rq\rq. The minimum communication performance requirement $R_c$ is set as 150 bits and the communication SNR is set as 10 dB. It is seen that the transmission energy increases with the minimum required MI. The reason is that the harvesting energy needs to be increased in order to satisfy the MI requirement. It is also seen that the EQ scheme requires a larger transmission energy than our proposed scheme. The reason is that our proposed scheme jointly optimizes all the variables related to the wireless resource. When the minimum required MI is lower than 220 bits and the radar SNR is 15dB, it is noted that the transmission energy required for the EQ scheme is almost constant. The reason is that the minimum MI achieved by the EQ scheme is 216 bits when the radar SNR is 15dB.

Fig. 2(b) shows the WPT time slot $\tau_1$ and IRC time slot $\tau_2$ versus the minimum required MI under different radar SNR and power allocation schemes. It is seen that, with the increase of the minimum required MI, the WPT time slot $\tau_1$ increases but the IRC time slot $\tau_2$ decreases. When the SNR is the same, the equal scheme requires a larger energy harvesting time $\tau_1$ than the optimal scheme and $\tau_1$ determines the upper bound of the system's performance. Therefore, given the total time and transmission power, our proposed  optimal scheme can achieve a greater MI than the equal power allocation scheme.

Fig. 2(c) shows the minimum transmission energy of the wireless power station versus the minimum required DIR. The minimum communication performance requirement $R_r$ is set as 150 bits and the radar SNR is set as 10 dB. It is seen that the transmission energy increases with the minimum required DIR, irrespective of the power allocation schemes. When the required DIR is lower than 160 bits and the communication SNR is 12dB, the transmission energy required for the OP scheme is almost constant. The reason is that the minimum DIR achieved by the OP scheme is 121 bits since $R_r$ is set as 150 bits. It is also seen that our proposed scheme outperforms the benchmark scheme in terms of energy consumption.

\section{Conclusion}



A wireless powered OFDM IRC system was studied. The optimal energy beamforming and waveform scheme was designed for minimizing the consumed energy. It was proved that the rank-one solutions can be obtained. Simulation results have verified the efficiency of our proposed scheme.

\end{document}